\documentclass[aps,pra,reprint,superscriptaddress,floatfix]{revtex4-2}
\usepackage[latin9]{inputenc}
\usepackage{bm}
\usepackage{amsmath}
\usepackage{amssymb}
\usepackage{graphicx}

\makeatletter

\newcommand{\lyxmathsym}[1]{\ifmmode\begingroup\def\b@ld{bold}
  \text{\ifx\math@version\b@ld\bfseries\fi#1}\endgroup\else#1\fi}

\usepackage{bm}
\usepackage{amsfonts}
\usepackage{dcolumn}
\usepackage{bm}

\makeatother

\begin{document}
\title{Disorder-dependent slopes of the upper critical field in nodal and
nodeless superconductors}
\author{V. G. Kogan}
\email{kogan@ameslab.gov}

\affiliation{Ames National Laboratory, Ames, Iowa 50011, U.S.A.}
\author{R. Prozorov}
\email{prozorov@ameslab.gov}

\affiliation{Ames National Laboratory, Ames, Iowa 50011, U.S.A.}
\affiliation{Department of Physics \& Astronomy, Iowa State University, Ames, IA
50011, U.S.A.}
\begin{abstract}
We study the slopes of the upper critical field $\partial_{T}H_{c2}|_{T_{c}}\equiv\partial H_{_{c2}}/\partial T$
at $T_{c}$ in anisotropic superconductors with transport (non-magnetic)
scattering employing the Ginzburg-Landau theory, developed for this
situation by S. Pokrovsky and V. Pokrovsky, Phys. Rev. B, \textbf{54},
13275 (1996). We found unexpected behavior of the slopes for a $d-$wave
superconductor and in a more general case of materials with line nodes
in the order parameter. Specifically, the presence of line nodes causes
$\partial_{T}H_{c2}|_{T_{c}}$ to \emph{decrease} with increasing
non-magnetic scattering parameter $P$, unlike the nodeless case where
the slope \emph{increases}. In a pure $d-$wave case, the slope $\partial H_{c2}|_{T_{c}}$
changes from decreasing to increasing when scattering parameter approaches
$P\approx0.91\,P_{{\rm crit}}$, where $P_{{\rm crit}}\approx0.2807$
at which $T_{c}\to0$ that implies the the existence of a gapless
state in $d-$wave superconductors with transport scattering in the
interval, $0.91\,P_{{\rm {crit}}}<P<P_{{\rm crit}}$. Furthermore,
we have considered the mixed $s+d$ order parameter that has 4 nodes
on a cylindrical Fermi surface when a $d-$part is dominant, or no
nodes at all when an $s-$phase is the major one. We find that presence
of nodes causes the slope $\partial_{T}H_{c2}|_{T_{c}},$ to decrease
initially with increasing $P$, whereas in the nodeless state, $\partial_{T}H_{c2}|_{T_{c}}$
monotonically increases. Therefore, fairly straightforward experiments
make it possible to decide whether or not the order parameter of a
superconductor has nodes by measuring the disorder-dependence of the
slope of $H_{c2}$ at $T_{c}$. 
\end{abstract}
\date{\today}
\maketitle

\section{Introduction}

A brief survey of the literature finds many reports on the upper critical
field slope at $T_{c}$ as a function of disorder in various materials.
The consistent picture emerges - superconductors with line nodes show
decreasing slope whereas those without nodes show increasing $\partial H_{c2}/\partial T$
with increasing disorder. For example, the increase was experimentally
observed in fully gapped $s_{\pm}$ iron-based superconductors where
nonmagnetic disorder was introduced by ball-milling \cite{Tokuta2019},
irradiation with 2.5 MeV electrons \cite{Prozorov2023} or fast neutrons
\cite{Karkin2014}. On the other hand, a decrease of the slope concomitant
with the decrease of $T_{c}$ was found in a nodal pnictide superconductor
\cite{Hashimoto2012}.

The problem of slopes $S=\partial H_{c2}/\partial T$ at $T_{c}$
(we also write it as, $\partial_{T}H_{c2}|_{T_{c}}$) can be addressed
with the help of Ginzburg-Landau (GL) theory. To provide a reasonable
theoretical guidance, the theory should be applicable to various order
parameter symmetries and to anisotropic Fermi surfaces. There were
a few attempts to develop such a version of GL, e.g. \cite{PS1,PS2}
have confirmed major features of the experimental information on slopes
dependence on disorder for superconductors with nodes and without.
However, some recent findings \cite{Konczykovski2023} of non-monotonic
disorder dependence of the slopes are puzzling. In this work we employ
a most general version of the GL theory applicable to anisotropic
order parameters in the presence of transport scattering due to S.
Pokrovsky and V. Pokrovsky, \cite{PP}.

Below, we first outline the theory, then we apply it to the d-wave
symmetry of the order parameter and show that, in fact, the slopes
$S$ are indeed suppressed by a relatively weak disorder but increase
when the impurity scattering approaches the critical value at which
$T_{c}=0$. Next, we consider the order parameter which is a superposition
of $s$ and $d$ that allows us to study slopes $S$ in cases of nodes
present or not and show that the nodes change $S(P)$ from the node-free
increase to decrease. We end up with a short discussion of existing
and possible experiments.

To simplify the formalism, the effective coupling is commonly assumed
factorizable \cite{Kad}, $V({\bm{k}},{\bm{k}}^{\prime})=V_{0}\,\Omega({\bm{k}})\,\Omega({\bm{k}}^{\prime})$,
$\bm{k}$ is the Fermi momentum. One then looks for the order parameter
in the form: 
\begin{equation}
\Delta({\bm{r}},T;{\bm{k}})=\Psi({\bm{r}},T)\,\Omega({\bm{k}})\,.\label{D=00003D00003DPsiO}
\end{equation}
The factor $\Omega({\bm{k}})$ for the order parameter change along
the Fermi surface is conveniently normalized: 
\begin{equation}
\langle\Omega^{2}\rangle=1\,,\label{norm}
\end{equation}
where $\langle...\rangle$ stand for the Fermi surface average. This
normalization corresponds to the critical temperature $T_{c0}$ of
a clean material given by the standard isotropic weak-coupling model
with the effective interaction $V_{0}$.

The slope of the upper critical field $H_{c2}$ at $T_{c}$ is determined
by the $T$-dependence of the coherence length $\xi=\xi_{GL}/\sqrt{1-T/T_{c}}$:
\begin{eqnarray}
\frac{\partial H_{c2}}{\partial T}\Big|_{T_{c}}=-\frac{\phi_{0}}{2\pi\xi_{GL}^{2}T_{c}}\,.\label{slp}
\end{eqnarray}
The length $\xi_{GL}$ is of the order of the BCS coherence length
$\xi_{0}=\hbar v_{F}/\pi\Delta(0)$, but differs from $\xi_{0}$,
actually it depends on the coupling, on the impurity scattering, on
the order parameter symmetry, and the Fermi surface anisotropy. All
these dependencies can, in principle, be found within the microscopic
BCS theory. This has been done at early stages of the theory of superconductivity
for anisotropic order parameter only in the clean limit by Gor'kov
and Melik-Barkhudarov \cite{Kad}, 
\begin{equation}
(\xi_{GL}^{2})_{ik}=\frac{7\zeta(3)\hbar^{2}}{16\pi^{2}T_{c}^{2}}\,\langle\Omega^{2}v_{i}v_{k}\rangle\,.
\label{xi-Gork}
\end{equation}

The transport scattering was included by Helfand and Werthamer \cite{HW}
but only for the isotropic order parameter and Fermi surface: 
\begin{equation}
\xi_{GL}^{2}=\frac{\hbar^{2}v^{2}}{24\pi^{3}T_{c}^{2}P^{2}}\left[\frac{\pi^{2}P}{4}-\psi\left(\frac{1+P}{2}\right)+\psi\left(\frac{1}{2}\right)\right]\,.\label{xi-HW}
\end{equation}
Here, $\psi$ is the di-gamma function and the scattering parameter
\begin{equation}
P=\hbar/2\pi T_{c0}\tau\label{P}
\end{equation}
with $\tau$ being the scattering time and $T_{c0}$ the critical
temperature for clean sample. It is easy to see that the slope $S=-\partial_{T}H_{c2}|_{T_{c}}$
grows being roughly proportional to $P$. It is worth paying attention
that the material parameter $P$ does not depend on actual $T_{c}$
unlike the scattering parameter $\rho=\hbar/2\pi T_{c}\tau$ often
employed in literature.

The critical temperature of materials with anisotropic order parameter
is suppressed even by non-magnetic impurities. This was established
by Openov \cite{Openov}, see also \cite{K2009}: 
\begin{eqnarray}
-\ln\frac{T_{c}}{T_{c0}}=(1-\langle\Omega\rangle^{2})\left[\psi\left(\frac{P/t_{c}+1}{2}\right)-\psi\left(\frac{1}{2}\right)\right],\qquad\label{tc}
\end{eqnarray}
where $t_{c}=T_{c}/T_{c0}$.

\begin{figure}[h]
\includegraphics[width=8cm]{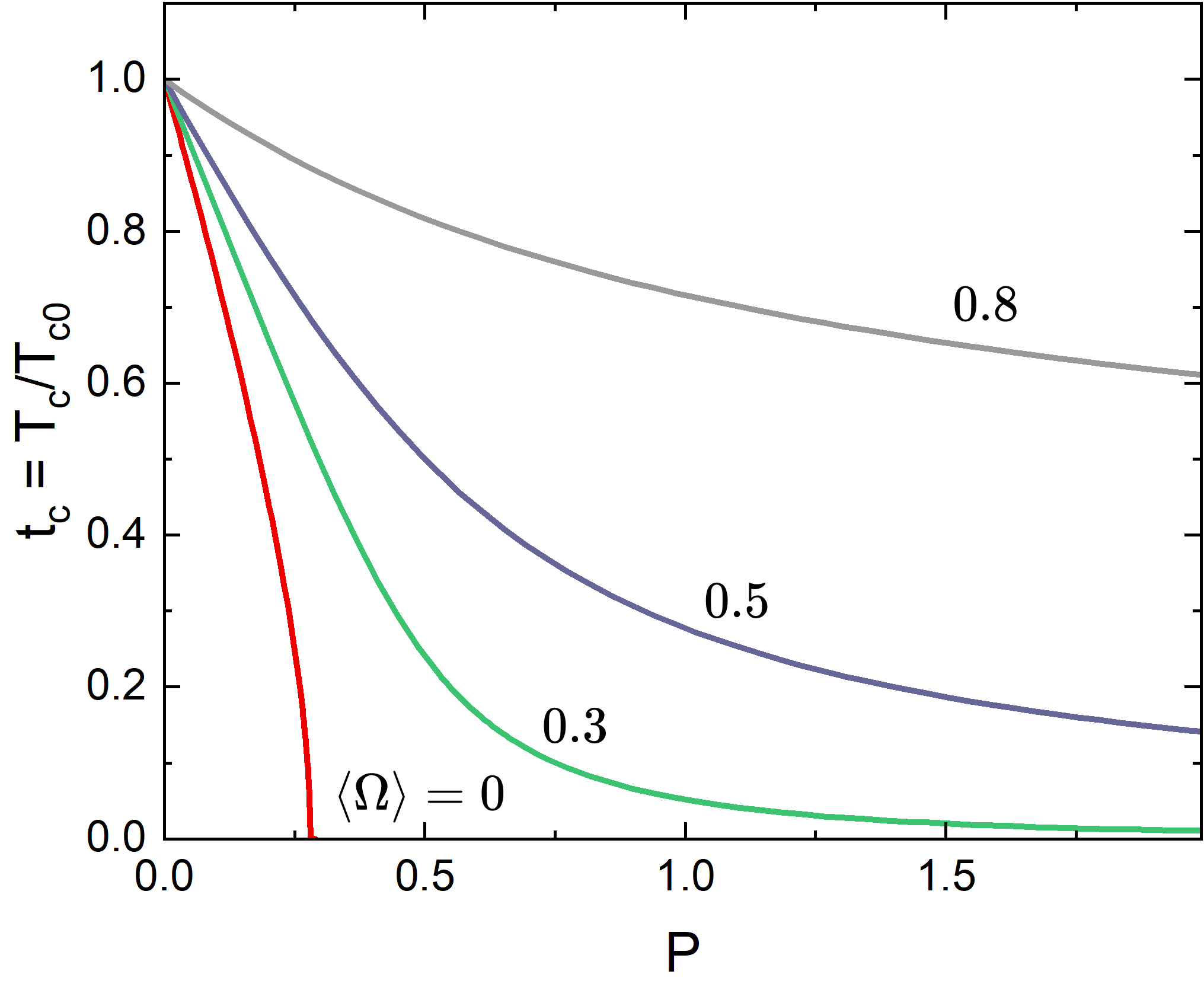} \caption{(Color online) $t_{c}=T_{c}/T_{c0}$ vs scattering parameter $P$
for $\langle\Omega\rangle=0.8,0.5,0.3,0$ in top-down order. Note
that for d-wave $\langle\Omega\rangle=0$ and $T_{c}$ turns zero
at the critical scattering $P_{crit}=1/2e^{\gamma}=0.2807$.}
\label{fig1} 
\end{figure}

Clearly, $T_{c}=T_{c0}$ for the isotropic order parameter with $\Omega=1$
and arbitrary scattering rate $P$ as well as for the clean limit
and any anisotropic $\Omega$. For the d-wave $\langle\Omega\rangle=0$
and we have the standard Abrikosov-Gor'kov result \cite{AG} so that
the transport scattering in d-wave affects $T_{c}$ as the pair-breaking
scattering in isotropic materials does.

\section{Linearized Ginzburg-Landau equations}

Thus, according to Eq.\,(\ref{slp}), the slope of $H_{c2}$ at $T_{c}$
is determined by the coherence length which enters linearized GL equation
for the order parameter $\Psi$, 
\begin{equation}
-(\xi^{2})_{ik}\Pi_{i}\Pi_{k}\Psi=\Psi\label{GL-linear}
\end{equation}
with the tensor of squared coherence length $(\xi^{2})_{ik}$; ${\bm{\Pi}}=\nabla+2\pi i{\bm{A}}/\phi_{0}$,
$\bm{A}$ is the vector potential and $\phi_{0}$ is the flux quantum.
For arbitrary $\Omega$ and any Fermi surface in presence of transport
scattering, A. Pokrovsky and V. Pokrovsky \cite{PP} evaluated the
tensor 
\begin{equation}
(\xi^{2})_{ik}=\zeta_{ik}/a\,.\label{PP1}
\end{equation}
Here, (in our notation) 
\begin{eqnarray}
a=\frac{T_{c}-T}{T_{c}}\left[1+(1-\langle\Omega\rangle^{2})\psi^{\prime}\left(\frac{P/t_{c}+1}{2}\right)\right]\,.\label{PP2}
\end{eqnarray}
The tensor $\hat{\zeta}$ is given by 
\begin{eqnarray}
\zeta_{ik} & = & \frac{\hbar^{2}}{16\pi^{2}T_{c}^{2}}\Big[h_{3,0}\langle\Omega^{2}v_{i}v_{k}\rangle\nonumber \\
 & + & \frac{P}{t_{c}}h_{3,1}\langle\Omega v_{i}v_{k}\rangle\langle\Omega\rangle+\frac{P^{2}}{4t_{c}^{2}}h_{3,2}\langle v_{i}v_{k}\rangle\langle\Omega\rangle^{2}\Big]\,.\label{PP3}
\end{eqnarray}
The quantities $h_{\mu,\nu}$ are functions of $x=P/2t_{c}$ defined
as 
\begin{eqnarray}
h_{\mu,\nu}(x)=\sum_{n=0}^{\infty}(n+1/2+x)^{-\mu}(n+1/2)^{-\nu}\,,\label{PP4}
\end{eqnarray}
so that 
\begin{eqnarray}
h_{3,0} & = & -\frac{1}{2}\psi^{\prime\prime}\left(\frac{1}{2}+x\right)\,,\nonumber \\
h_{3,1} & = & \frac{1}{x^{3}}\Big[\psi\left(\frac{1}{2}+x\right)-\psi\left(\frac{1}{2}\right)-x\psi^{\prime}\left(\frac{1}{2}+x\right)\nonumber \\
 & + & \frac{x^{2}}{2}\psi^{\prime\prime}\left(\frac{1}{2}+x\right)\Big]\,,\nonumber \\
h_{3,2} & = & \frac{1}{2x^{4}}\Big\{\pi^{2}x-6\left[\psi\left(\frac{1}{2}+x\right)-\psi\left(\frac{1}{2}\right)\right]\nonumber \\
 & + & 4x\psi^{\prime}\left(\frac{1}{2}+x\right)-x^{2}\psi^{\prime\prime}\left(\frac{1}{2}+x\right)\Big\}.\label{PP5}
\end{eqnarray}

It is straightforward to see that for the isotropic order parameter
with $\Omega=1$ on the Fermi sphere these formulas reduce to the
BCS form 
\begin{eqnarray}
\xi^{2}=\frac{7\zeta(3)}{48\pi^{2}T_{c}^{2}\delta t}\chi(P)\,,\quad\delta t=1-\frac{T}{T_{c}}\,,\label{PP6}
\end{eqnarray}
with the Gor'kov function 
\begin{eqnarray}
\chi(P)=\frac{1}{7\zeta(3)}\sum_{n=0}^{\infty}\frac{1}{(n+1/2+P)(n+1/2)^{2}}\,.\label{PP7}
\end{eqnarray}
This limit can also be checked by comparison with $H_{c2}$ slopes
at $T_{c}$ given by Helfand and Werthamer \cite{HW}.

\begin{figure}[h]
\includegraphics[width=8cm]{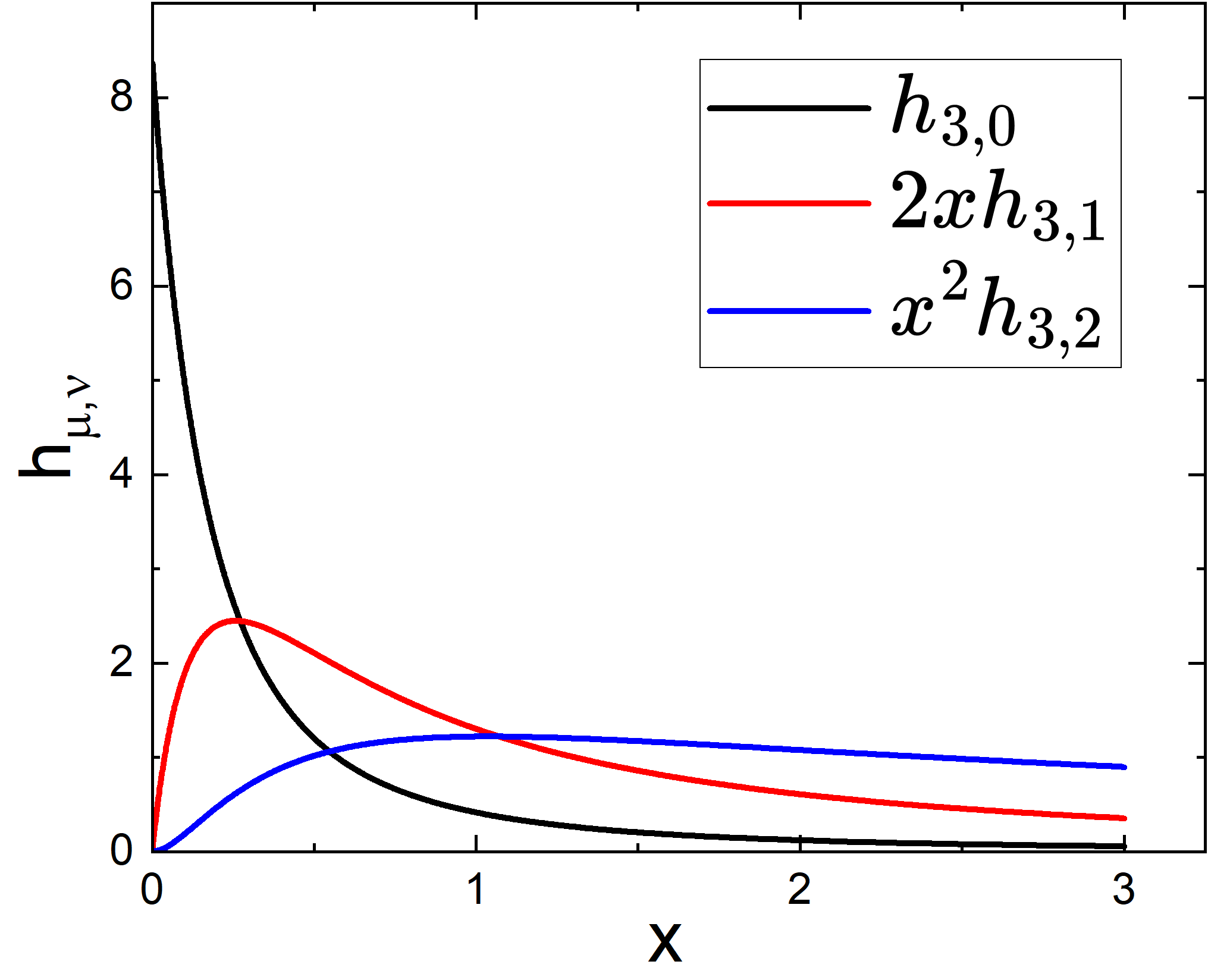} \caption{(Color online) The coefficients $h_{3,0}(x)$, $2xh_{3,1}(x)$, and
$x^{2}h_{3,2}(x)$ vs $x$.}
\label{f2} 
\end{figure}

It is worth noting that for the clean limit $x=P/2t_{c}\to0$, and
$h_{3,0}\to7\zeta(3)$, $h_{3,1}\to\pi^{4}/6$, and $h_{3,2}\to31\zeta(5)$.
This means that in this limit, only the first term on RHS of Eq.\,(\ref{PP3})
survives in agreement with \cite{Kad}. In the opposite limit $x=P/2t_{c}\gg1$,
the last term in Eq.\,(\ref{PP3}) dominates \cite{PP} as is seen
in Fig.\,\ref{f2} (however, this limit does not apply for the d-wave
since both the second and the third terms in Eq.\,(\ref{PP3}) are
zeroes due to $\langle\Omega\rangle=0$). 

\subsection{General order parameter }
\begin{widetext}
In he general case we obtain for the slope of $H_{c2}$ along the
$c$ axis of a uniaxial material 
\begin{eqnarray}
\frac{\partial H_{c2}}{\partial T}\Big|_{T_{c}}=-\frac{8\pi\phi_{0}T_{c0}}{\hbar^{2}}\frac{t_{c}\left[1+(1-\langle\Omega\rangle^{2})\psi^{\prime}\left(\frac{1+P/t_{c}}{2}\right)\right]}{h_{3,0}\langle\Omega^{2}v_{a}^{2}\rangle+2(P/2t_{c})h_{3,1}\langle\Omega\rangle\langle\Omega v_{a}^{2}\rangle+(P/2t_{c})^{2}h_{3,2}\langle\Omega\rangle^{2}\langle v_{a}^{2}\rangle}\,.\qquad\label{general}
\end{eqnarray}
\end{widetext}

Here, all coefficients $h_{\mu,\nu}(x)$ are taken at $x=P/2t_{c}$.

Since the Fermi velocity is not a constant at anisotropic Fermi surfaces,
we normalize velocities on some value $v_{0}$ for which we choose
\cite{KP} 
\begin{equation}
v_{0}^{3}=2E_{F}^{2}/\pi^{2}\hbar^{3}N(0)\,,\label{v0}
\end{equation}
where $E_{F}$ is the Fermi energy and $N(0)$ is the total density
of states at the Fermi level per spin. One easily verifies that $v_{0}=v_{F}$
for the isotropic case.

The slope expression (\ref{general}) remains the same except a changed
pre-factor 
\begin{equation}
-\frac{8\pi\phi_{0}T_{c0}}{\hbar^{2}}\to-\frac{8\pi\phi_{0}T_{c0}}{\hbar^{2}v_{0}^{2}}\,,\label{prefactor}
\end{equation}
and the velocity $v_{a}$ is now dimensionless (although we leave
for it the same notation).

\subsection{d-wave}


The case of the d-wave symmetry of the order parameter with $\langle\Omega\rangle=0$
is relatively simple. We have the coherence length relevant for $H_{c2}$
along $c$ axis of a uniaxial crystal: 
\begin{eqnarray}
\xi_{aa}^{2} & = & \frac{\zeta_{aa}}{a}=\frac{\hbar^{2}\langle\Omega^{2}v_{a}^{2}\rangle h_{3,0}}{16\pi^{2}T_{c}^{2}\left[1+\psi^{\prime}(P/2t_{c}+1/2)\right]\delta t}.\label{xi_aa}
\end{eqnarray}

For a Fermi cylinder, with $\Omega=\sqrt{2}\cos2\varphi$, the average
$\langle\Omega^{2}v_{a}^{2}\rangle=v^{2}/2$. Hence, 
\begin{eqnarray}
\xi_{aa}^{2}=-\frac{\hbar^{2}v^{2}\,\psi^{\prime\prime}(P/2t_{c}+1/2)}{64\pi^{2}T_{c}^{2}\left[1+\psi^{\prime}(P/2t_{c}+1/2)\right]\delta t}.\label{xi_aa2}
\end{eqnarray}
and we obtain: 
\begin{eqnarray}
\frac{dH_{c2,c}}{dT}\Big|_{T_{c}}=\frac{32\pi\phi_{0}T_{c0}}{\hbar^{2}v^{2}}\,t_{c}\frac{1+\psi^{\prime}(P/2t_{c}+1/2)}{\psi^{\prime\prime}(P/2t_{c}+1/2)}.\label{slpd}
\end{eqnarray}

For numerical work it is convenient to use the reduced slope 
\begin{equation}
S=-\frac{dh_{c2}}{dt}\Big|_{t_{c}}=-\frac{\hbar^{2}v^{2}}{8\pi\phi_{0}T_{c}}\frac{dH_{c2}}{dT}\Big|_{T_{c}}\,.\label{eq:reduced}
\end{equation}
While the actual slope $\left.\partial H_{c2}/\partial T\right|_{T_{c}}$
is negative, we are interested in its magnitude and use a positive
quantity as given by Eq.\,(\ref{eq:reduced}).

\begin{figure}[h]
\includegraphics[width=8cm]{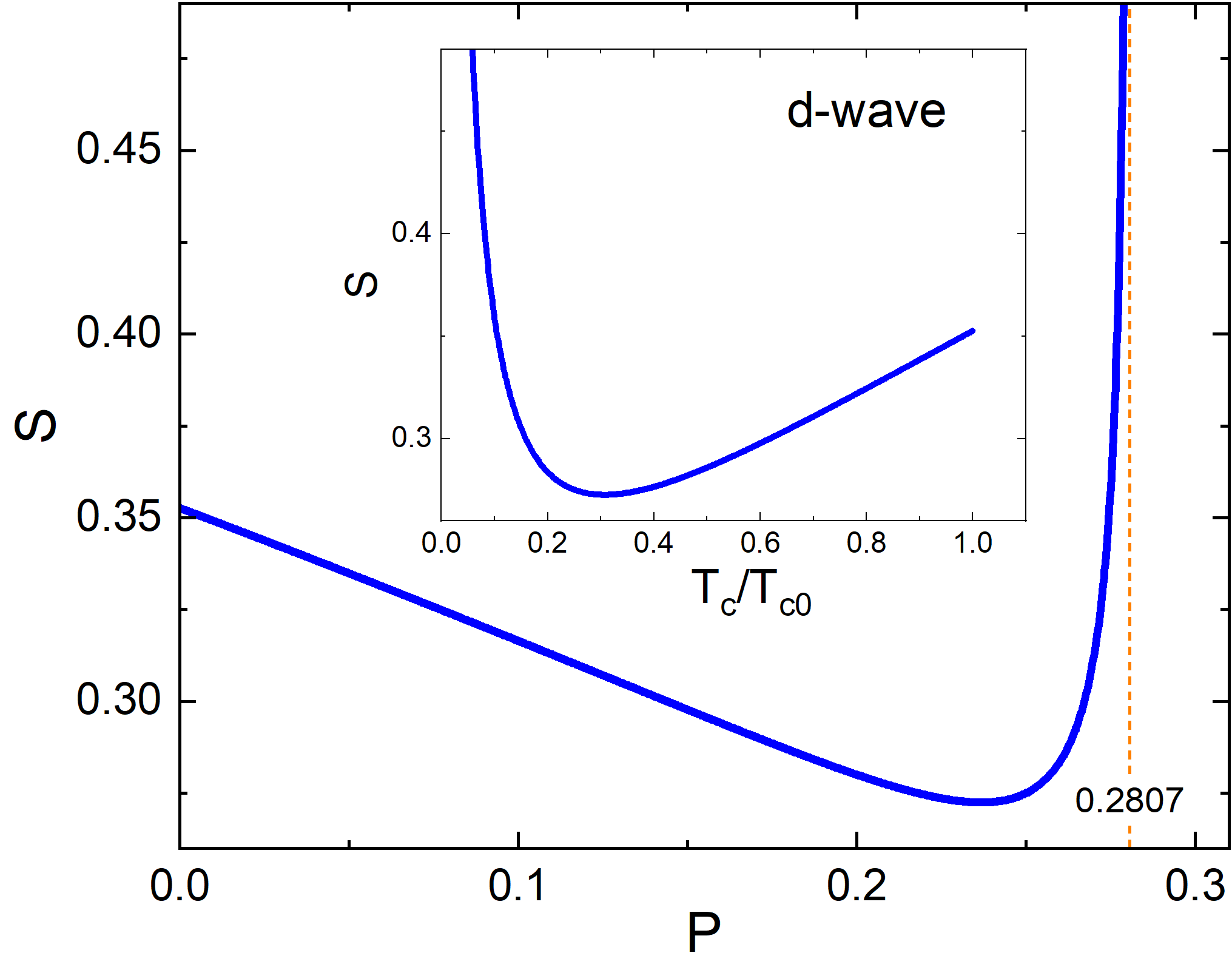} \caption{(Color online) The slope $S=\partial h_{c2}/\partial h$ at $t_{c}$
according to Eq.\,(\ref{eq:reduced}) vs $P$ for a d-wave material.
Inset: the same slopes vs $t_{c}=T_{c}/T_{c0}$. 
}
\label{f3} 
\end{figure}

The behavior of the slope as a function of $P$ according to this
result is shown in Fig.\,\ref{f3}. As is known, the maximum scattering
parameter $P$ for which the d-wave superconductivity survives is
$P_{{\rm {crit}}}=1/2e^{\gamma}=0.28$ (the double of the critical
value for the spin-flip magnetic scattering \cite{AG}). Hence, similar
to materials with magnetic scatterers \cite{KP2013}, the slope $\partial H_{c2}/\partial T$
at $T_{c}$ for d-wave decreases with increasing transport scattering.
It is worth noting that this behavior changes to increase near $P\approx0.91\,P_{{\rm {crit}}}=0.25$;
this estimate coincides with that given by Abrikosov and Gor'kov for
the low bound of the gapless state. This suggests that d-wave materials
can also be gapless if the transport scattering parameter lays in
the interval $0.25<P<0.28$. Thus, $S(P)$ dependence might be a macroscopic
manifestation of the gapless superconductivity in impure d-wave materials,
the speculation worth of further study.

Figure \ref{f4} shows the slopes of Fig.\,\ref{f3} plotted vs the
critical temperate $t_{c}(P)$.

\subsection{$\bm{s+d}$}

Obviously the major interest in the community is to determine whether
the easily accessible measurements of the $H_{c2}$ slope near $T_{c}$,
may provide some insight into the structure of the order parameter.
Here we examine the simplest case of a $s+d$ state where order parameter
is the isotropic s-wave in one limit and a standard 2D $d-$wave in
the other. Keeping in mind the normalization, $\langle\Omega^{2}\rangle=1,$
a convenient order parameter can be written as: 
\begin{equation}
\Omega=\sqrt{1-r^{2}/2}+r\cos(2\varphi)\label{eq:s+d}
\end{equation}
If $r=0$, $\Omega=1$ and when $r=\sqrt{2}$, $\Omega=\sqrt{2}\cos(2\varphi)$
which are the required limiting cases. We choose this order parameter
not because it may describe any particular real material, rather we
intend to check whether or not there is a connection between the microscopic
anisotropy of the order parameter and the macroscopic dependence $S(P)$
of slopes of $H_{c2}$ on the degree of disorder $P$.

\begin{figure}[ht]
\includegraphics[width=8cm]{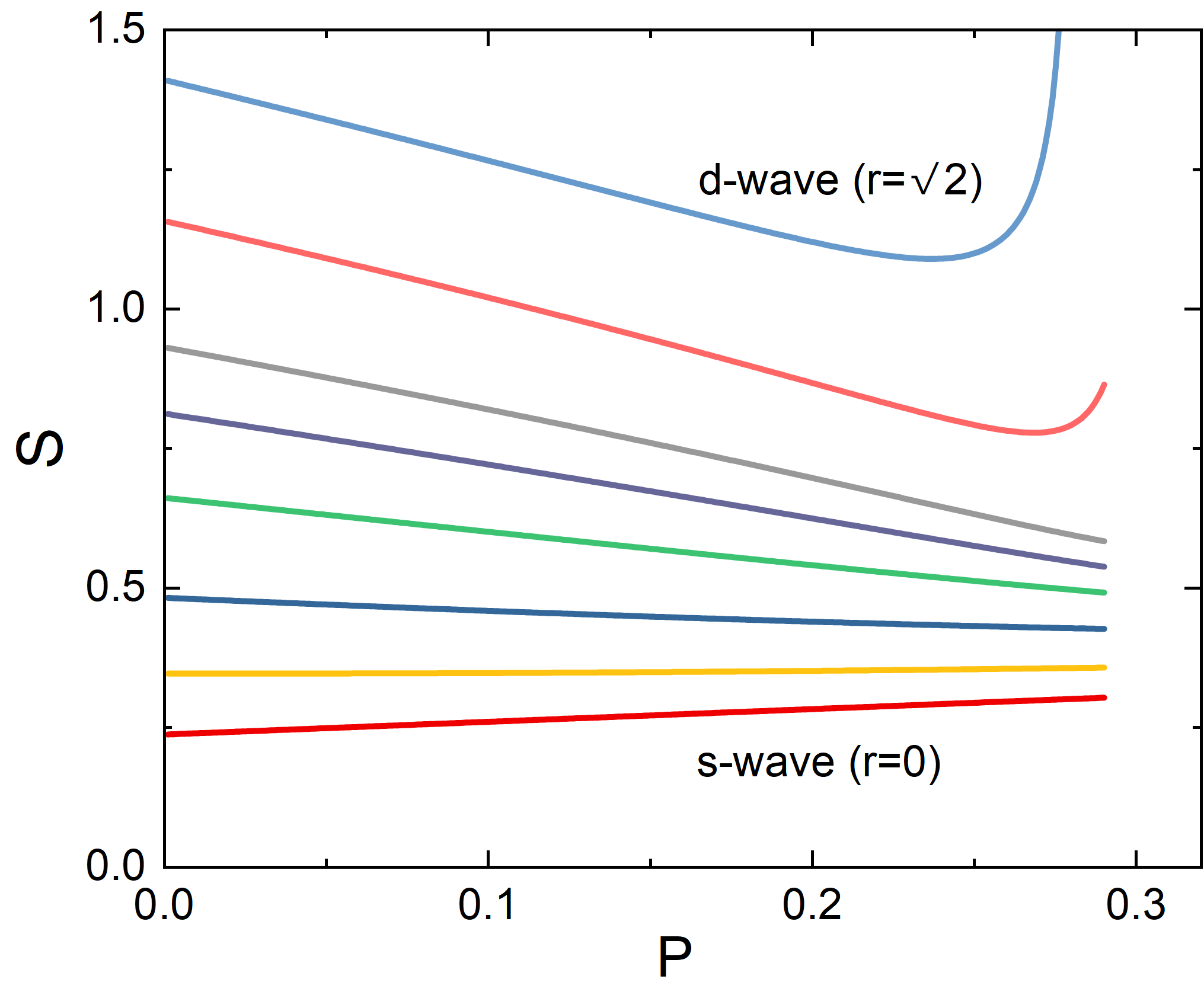} \caption{(Color online) $S(P)$ according to Eqs.\,(\ref{eq:reduced}) and
(\ref{general}) for a set of parameter $r=0,0.75,1,1.2,1.3,1.35,1.4$
and $\sqrt{2}$. The approach to a gapless regime is characterized
by an upturn starting with $P\approx0.25$ for a $d$-wave order parameter
with $r=\sqrt{2}$. The calculations were performed for a cylindrical
Fermi surface.}
\label{f4} 
\end{figure}

\begin{figure}
\includegraphics[width=8cm]{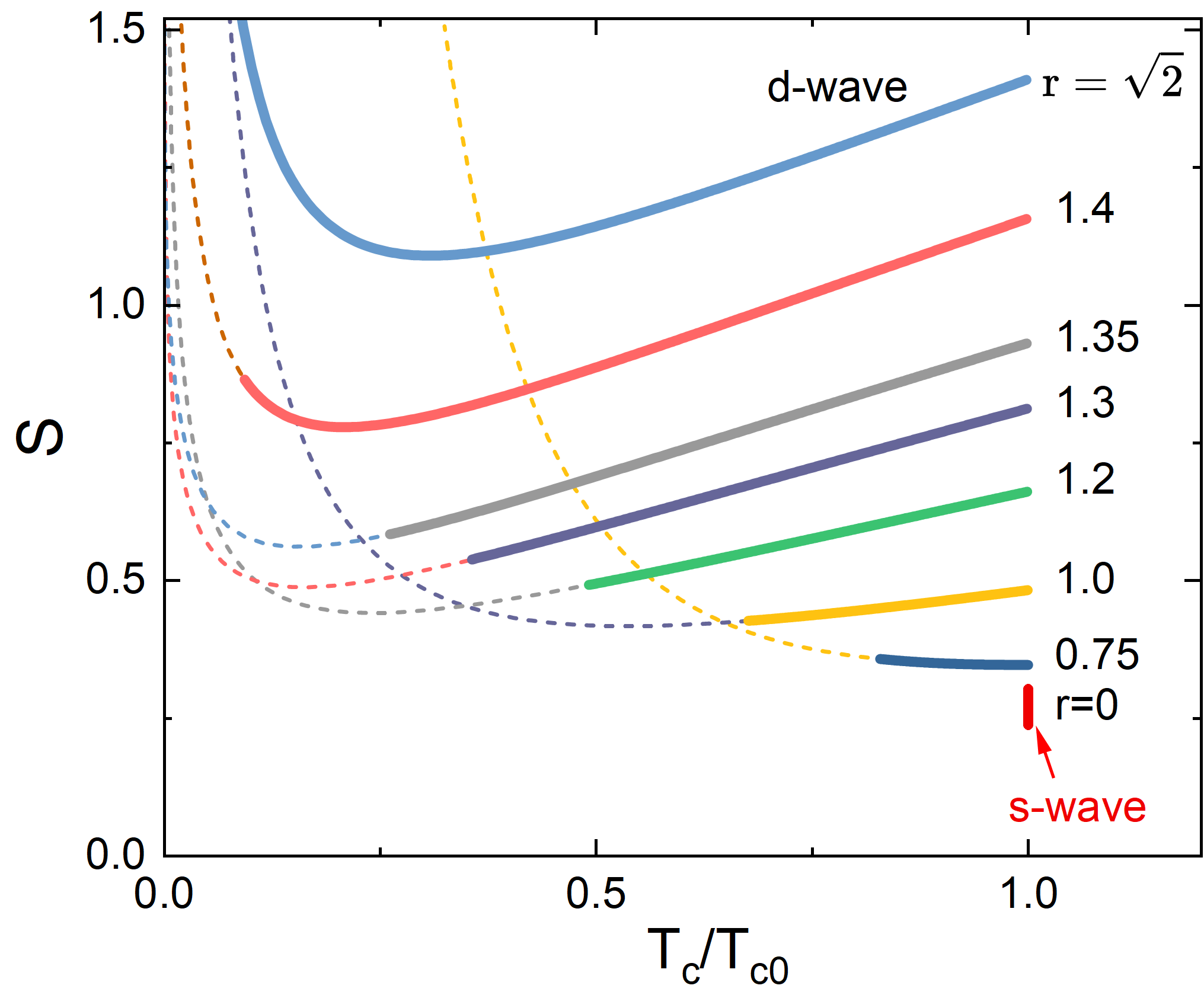} \caption{(Color online) Solid lines are the slopes $S$ for the same values
of the scattering parameters $P$ as in Fig.\,\ref{f4}, but plotted
versus the transition temperature $t_{c}(P)$. Dashed continuations
are for $P>0.28$ for which the pure d-wave phase does not exist.
In the pure $s$-wave for which $t_{c}=1$, $S(P)$ does not depend
on $P$ and shown by a vertical red line.}
\label{f5} 
\end{figure}

\begin{figure}
\includegraphics[width=8cm]{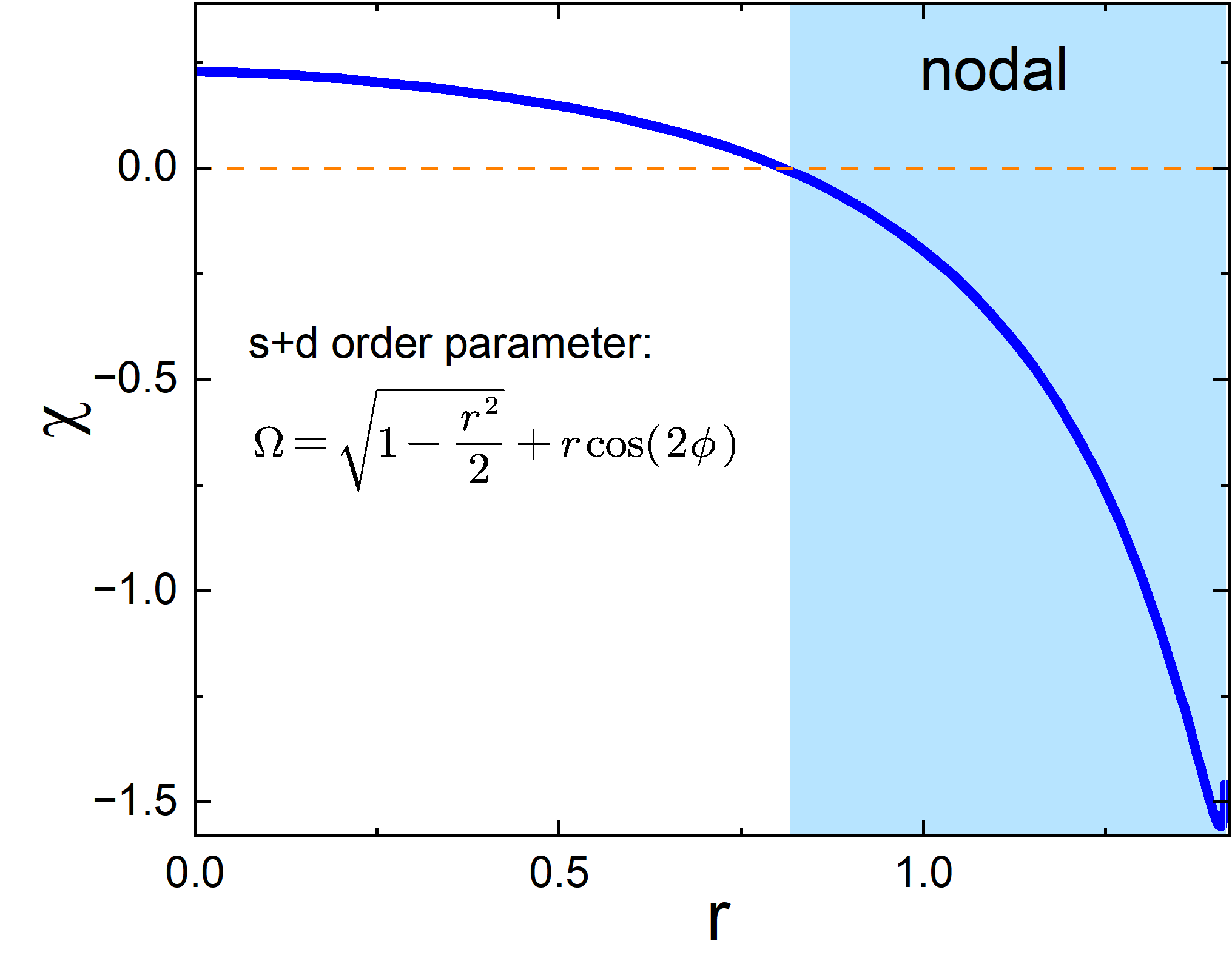} \caption{(Color online) The rate of change of the slope $S(P)$ defined as
$\chi=S(0.11)-S(0.01))/0.1$, evaluated for different values of the
$r$-coefficient. 
Nodal behavior starts at $r=\sqrt{2/3}\approx0.817$. Apparently,
this coincides with the rate of slope change becoming negative.}
\label{f6} 
\end{figure}

Figure \ref{f4} shows $S(P)$ calculated with the help of Eqs.\,(\ref{general}),
(\ref{eq:reduced}), plotted for a few coefficients $r$ of the $\Omega$
function, Eq.\,(\ref{eq:s+d}). At $r=0$ the order parameter is
the isotropic $s$-wave, whereas at $r=\sqrt{2}$ it is a two-dimensional
d-wave order parameter. The calculations were performed for a cylindrical
Fermi surface keeping in mind possible applications of this work to
high-$T_{c}$ cuprates. Figure \ref{f4} shows that the slopes $S(P)$
for purely d-wave are (a) non-monotonic and decrease with increasing
$P$ up to about $P=0.25$ following by divergence when $P\to0.28$,
as we have seen in Fig.\,\ref{f3}. With increasing fraction of s-wave,
however, the negative slope of $S(P)$ for small and intermediate
$P$ weakens and turns to positive nearly linear increase of $S$.
To gain a further insight, we plot $S(P)$ versus $t_{c}$ in Fig.\,\ref{f5}. 

Since the overall behavior of the slopes $S(P)$ depends on fractions
of $s$ and $d$ phases, we consider the relative difference $\chi=S(0.11)-S(0.01))/0.1$
which characterizes this behavior, i.e. increase for $\chi>0$ and
decrease for $\chi<0$. Figure \ref{f6} shows this difference plotted
versus coefficient $r$ of the order parameter, Eq.\,(\ref{eq:s+d}).
A straightforward algebra shows that the order parameter is nodeless
(anisotropic $s$) for $r<\sqrt{2/3}\approx0.82$ and has four line
nodes for $r>\sqrt{2/3}$. Remarkably, Fig.\,\ref{f6} shows that
the rate of slope change $\chi$ becomes negative as soon as the nodes
appear. In other words, if one would measure $\left|S\right|=|\partial H_{c2}/\partial T|_{T_{c}}|$
as a function of non-magnetic disorder, the increasing $S$ would
indicate nodeless superconductivity, whereas the slope decrease can
be considered as evidence for the nodes presence.

\section{Discussion}

We have applied the GL theory developed by S. Pokrovsky and V. Pokrovsky
\cite{PP} for anisotropic materials in the presence of non-magnetic
scattering to evaluate slopes $\partial H_{c2}/\partial T|_{T_{c}}$
for d-wave superconductors and for the case of mixed $s+d$ symmetry.
For the d-wave, we find that for weak and intermediate scattering
rates $P=\hbar/2\pi T_{c0}\tau$, the slopes $S=-\partial H_{c2}/\partial T|_{T_{c}}$
are suppressed, similar to the situation of the magnetic pair-breaking
disorder and opposite to the transport scattering enhancement of slopes
$S(P)$ for the $s$-wave. One example of such a behavior is in studies
of slopes in YBaCuO by Antonov \textit{et al.} \cite{Antonov}.

Unexpectedly however, if the scattering rate approaches the critical
value $P_{crit}=0.28$ for which $T_{c}\to0$, the slopes increase
starting with $P\approx0.25$ to diverge at $P_{crit}$. The value
$0.25\approx0.9$ of $P_{crit}$, the fraction found by Abrikosov
and Gor'kov for the low bound of the gapless domain in standard $s$-wave
materials with magnetic impurities \cite{AG}. This analogy suggests
strongly that in d-wave with non-magnetic disorder close to the maximum
disorder possible we are dealing with a kind of ``gapless" state.
In our view, this speculation deserves careful examination.

The slopes at $T_{c}$ are the easiest thing one can measure after
a new superconductor is discovered. It has been done \textit{in-situ}
even for room-temperature materials under extremely high pressures
within diamond pressure cells. If indeed, the easily done macroscopic
measurement of slopes at $T_{c}$ may provide even partial information
on the microscopic symmetry of the order parameter, this would be
a worthy thing to do.

Thinking along these lines, we have examined the mixed $s+d$ order
parameter $\Omega(r)=\sqrt{1-r^{2}/2}+r\cos(2\varphi)$ such that
$\Omega(0)$ corresponds to the pure isotropic $s$-wave and $\Omega(\sqrt{2})$
is the pure $d$. We find the remarkable one-to-one correspondence
between presence or absence of nodes and decrease or increase of the
$H_{c2}$ slopes at $T_{c}$. Practically, this is a highly useful
observation, notwithstanding our oversimplified model.

There are various ways to introduce point-like uniform disorder into
superconductors. Perhaps the most promising and controllable are irradiation
with neutrons or electrons, see e.g. \cite{Karkin2014} and \cite{Konczykovski2023}.
To our knowledge, the slope $S$ in a nodal superconductor BaFe$_{2}$(As,P)$_{2}$
is decreasing up to large amounts of disorder \cite{Konczykovski2023}.

Hence, one of our major results is a novel experimentally-simple way
to distinguish between nodeless and nodal superconductivity. One has
to examine the rate of change of the slopes in samples of the same
chemical composition but with different amount of non-magnetic disorder.
Such disorder can be induced by ion implantation \cite{Wilke2004,Antonov},
electron \cite{Prozorov2023}, proton \cite{Kim2012}, neutron \cite{Karkin2014,Wilke2006},
or gamma \cite{Hamalawy1992} and even alpha particle irradiation
\cite{Tarantini2018}. If it decreases, the system is likely nodal.
If not, it is not.

A word of caution. It is quite possible that complex multi-band materials
will not follow our simple scheme, employing the generally accepted
factorization of temperature and angular variations of the order parameter.
It is also possible that other types of order parameters, say triplet,
will not follow either. However, our analysis is based on universal
Ginzburg-Landau theory and, hopefully, our conclusions may turn out
robust.
\begin{acknowledgments}
We thank M. Tanatar and S. Bud'ko for discussions. This work was supported
by the U.S. Department of Energy (DOE), Office of Science, Basic Energy
Sciences, Materials Science and Engineering Division. Ames Laboratory
is operated for the U.S. DOE by Iowa State University under contract
DE-AC02-07CH11358. 
\end{acknowledgments}

\bibliographystyle{apsrev4-2}

\begin{thebibliography}{10}
\bibitem{Konczykovski2023} A significant suppression of $T_{c}$
and of the slope of the upper critical field was observed in line-nodal
iron pnictide, BaFe$_{2}$(As,P)$_{2}$. {[}Private communication{]}.
\end{thebibliography}

\end{document}